\documentclass[twocolumn,amsmath,amssymb,footinbib,superscriptaddress]{revtex4-2}

\usepackage{graphicx,color}
\usepackage{dcolumn}
\usepackage{bm}
\usepackage{url}
\usepackage{amsmath}
\usepackage[colorlinks=true,citecolor=blue,urlcolor=blue,linkcolor=blue]{hyperref}
\usepackage[normalem]{ulem}
\usepackage{enumerate}

\begin{document}

\title{Suppression of shear ionic motions in bismuth\\ by coupling with large-amplitude internal displacement}

\author{Kunie Ishioka}
\email{ishioka.kunie@nims.go.jp}
\affiliation{National Institute for Materials Science, Tsukuba, 305-0047 Japan}

\author{Oleg V. Misochko}
\affiliation{Institute of Solid State Physics, Russian Academy of Sciences, 142432 Chernogolovka, Moscow region, Russia}

\date{\today}

\begin{abstract}

Bismuth, with its rhombohedral crystalline structure and two Raman active phonon modes corresponding to the internal displacement ($A_{1g}$) and shear ($E_{g}$) ionic motions, offers an ideal target for the investigation of the non-equilibrium phonon-phonon and electron-phonon couplings.  We investigate the $E_g$ phonon dynamics under intense photoexcitation by performing anisotropic transient reflectivity (TR) measurements on a 1-mm thick bismuth single crystal at 11 K.  The amplitude of coherent $E_g$ phonon is found to increase with incident pump fluence up to 10\,mJ/cm$^2$ and then turns to an apparent decrease.  This behavior is in stark contrast to the amplitude of the $A_{1g}$ phonon in standard TR measurements, which increases monotonically up to 20\,mJ/cm$^2$ and then saturates.  The contrasted behaviors of the two phonon modes can be interpreted in terms of the strong coupling of the $E_g$ oscillation with large-amplitude $A_{1g}$ displacement on a highly excited electronic state, where dynamic fluctuation of the vibrational potential would lead to a quick loss in the $E_g$ vibrational coherence.  Unlike  the previous studies on thin Bi films on substrates we observe no sign of a transition to a high-symmetry phase but a sign of partial damage of the crystalline surface at 28\,mJ/cm$^2$, possibly due to less efficient cooling at the surface of a bulk crystal.
 
\end{abstract}


\maketitle

\section{Introduction}

\begin{figure}
\includegraphics[width=0.475\textwidth]{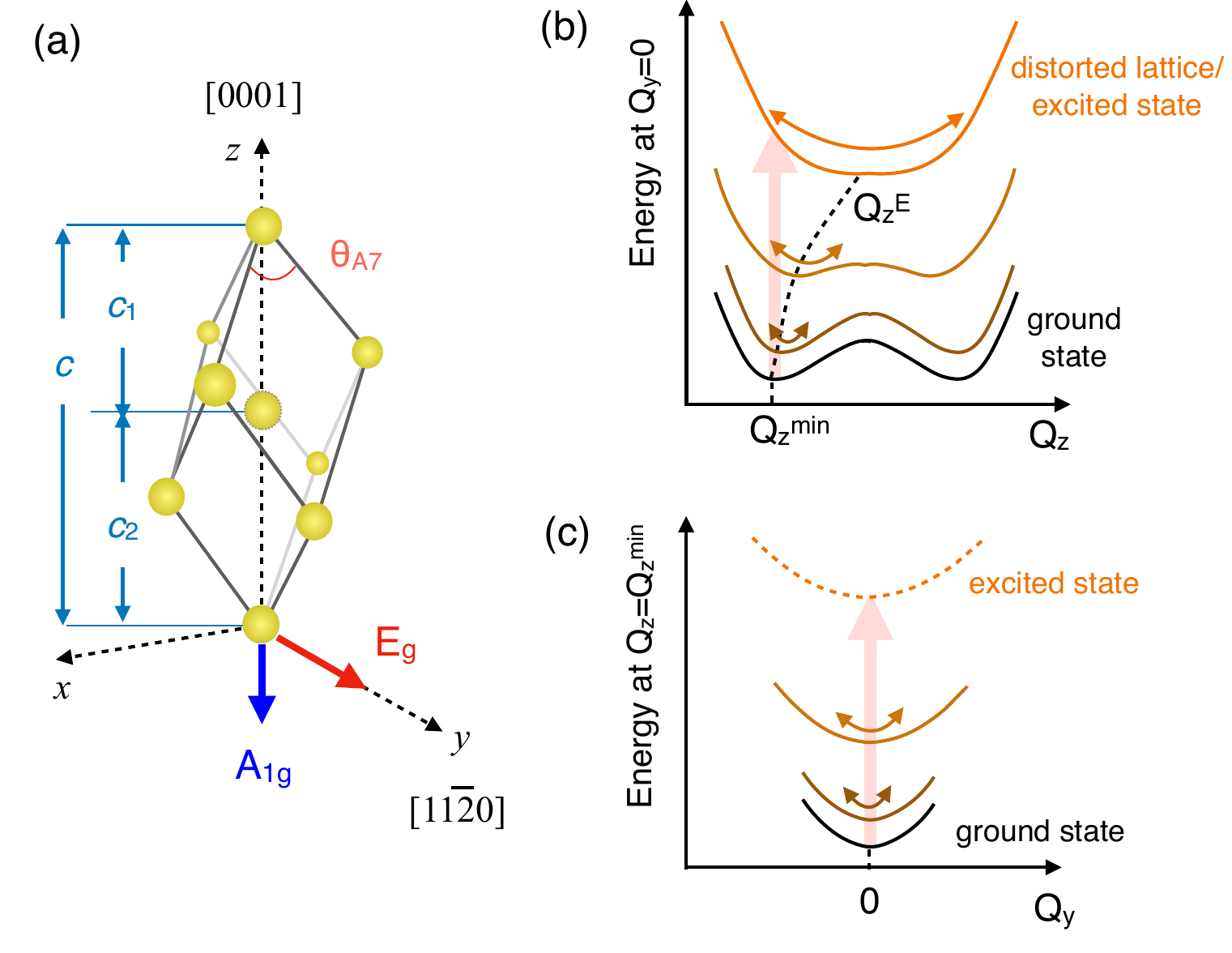}
\caption{\label{rhombo} (a) Crystalline structure of Bi together with the directions of the $A_{1g}$ and $E_g$ displacements. (b, c) Schematic illustrations of potential energy surfaces (PESs) along the $A_{1g}$ (b) and $E_g$ (c) coordinates for the ground state and for the lattice under photoexcitation or distortion.   
}
\end{figure}

Bismuth (Bi) is recently attracting renewed attention because of its surface accommodating topological insulating electronic states.  As for the bulk Bi electronic states, it is established that Bi is a semimetal with its Fermi surface consisting of three electron pockets at the $L$ points of the Brillouin zone and one hole pocket at the $T$ point  \cite{Shick1999}, though their topological classification is still under debate \cite{Ohtsubo2013, Aguilera2015, Schindler2018, Nayak2019, Jin2020}.  
The crystalline structure of Bi crystal is rhombohedral with $A7$ symmetry, as illustrated in Fig.~\ref{rhombo}(a), with the ground-state internal displacement along the trigonal ($z$) axis $u=c_1/2(c_1+c_2)=0.2357$ and the trigonal shear angle $\theta_{A7}=57.23^\circ$ \cite{Aguilera2015, Wu2018, Wu2019}.  This structure would be transformed to simple cubic by a slight deformation to $u=0.25$ and $\theta_{A7}=60^\circ$ \cite{Zouhar2016, Shick1999}.  Theoretical simulations predicted that the electronic band structure of Bi could undergo phase transitions from a semimetal to a semiconductor or to a metal by tweaking $u$ and/or $\theta_{A7}$ \cite{Shick1999, Aguilera2015, Wu2018, Wu2019}.  Correspondingly, a pressure-induced semimetal-semiconductor phase transition of Bi was reported experimentally \cite{Armitage2010, Brown2015}.

The rhombohedral Bi crystal features two Raman-active modes of $A_{1g}$- and $E_g$-symmetries, which are associated with the variations in $u$ and $\theta_{A7}$, respectively, as shown in Fig.~\ref{rhombo}(a).  Theoretical simulations predicted that introduction of photocarriers \cite{Murray2005, Zijlstra2006} transforms the potential energy surface (PES) along the $A_{1g}$ coordinate from a double well into a flattened single well, as schematically illustrated in Fig.~\ref{rhombo}(b).  
The susceptibility of the PES to the electronic excitation leads to a significant enhancement of coherent $A_{1g}$ phonons via ``displacive excitation of coherent phonons (DECP)" mechanism \cite{Cheng1991, Zeiger1992} and has thereby made the $A_{1g}$ mode a model target for optical studies on non-equilibrium electron-phonon coupling \cite{Cheng1990, Cheng1991, Hase2002, Misochko2004, Ishioka2006, Boschetto2008, Katsuki2013, Sheu2013, Misochko2014, Cheng2017, Geneaux2021, Thiemann2022, Papalazarou2012, Leuenberger2013, Perfetti2015}.  In the DECP generation mechanism a photoinduced shift in the equilibrium coordinate 
gives a driving force that is dependent on excited carrier density $N$ \cite{Zeiger1992}.  
If the carriers are created sufficiently fast and live sufficiently long (displacive limit), the coherent ionic displacement $Q_z$ can be described by a shifted damped harmonic function:
\begin{equation}\label{Eq2}
Q_z(t)=Q^E_z -Q_{z}^0\exp(-\Gamma_z t)\cos(2\pi\nu_z t),
\end{equation}
with the initial amplitude defined by the equilibrium shift: $Q_z^0(N)=Q_z^E(N)-Q_z^\text{min}$.  
The DECP enhancement of the $A_{1g}$ mode of Bi was experimentally confirmed by means of time-resolved x ray diffraction (trXRD) \cite{STinten2003, Fritz2007, Johnson2008, Johnson2013, Teitelbaum2021, Kubota2023} and was supported by density functional theory (DFT) calculations \cite{Murray2005, Zijlstra2006, Fritz2007, Giret2011}. 
One of the interests of recent ultrafast spectroscopic experiments on Bi lies on the optical control of its structural and electronic phases.  A single-shot transient reflectivity (TR) study on thin Bi films \cite{Teitelbaum2018} observed the disappearance of the coherent $A_{1g}$ oscillation signal under intense photoexcitation and attribute it to the theoretically predicted transition to a high-symmetry phase with a single-well PES.  The threshold fluence for the phase transition was found to depend critically on the Bi film thickness, suggesting that the phase transition competes with the electronic transport in the depth direction on sub-picosecond time scale \cite{Thiemann2022, Jnawali2021}.

The $E_g$-symmetry mode of Bi has been much less explored by time-resolved experiments, despite of the theoretical predictions that it could lead to a semiconductor-semimetal phase transition \cite{Shick1999, Aguilera2015, Wu2018}.  This is partly because of its much smaller amplitude compared to the $A_{1g}$ mode \cite{Misochko2007, Johnson2013} as a result of the insusceptibility of its equilibrium position to photoexcitation, as schematically illustrated in Fig.~\ref{rhombo}(c).  
The anisotropic optical polarization-dependence confirmed the generation of the $E_g$ phonon to be dominated by impulsive stimulated Raman scattering (ISRS) mechanism at low excitation densities \cite{Ishioka2006}. 
With a sufficiently short light field and a short-lived intermediate state (impulsive limit) the ionic displacement can be expressed by \cite{Dhar1994, Dekorsy}:
\begin{equation}\label{Eq4}
Q_y(t)=Q_y^0\exp(-\Gamma_y t)\sin(2\pi\nu_y t).
\end{equation}
A density functional theory (DFT) study \cite{Zijlstra2006} predicted a strong coupling between the $E_g$ and $A_{1g}$ modes under intense photoexcitation. 
The coupling was experimentally demonstrated as the emergence of a combination mode at the difference frequency in a prior TR study \cite{Misochko2006}.  The literature also reported the $E_g$ and $A_{1g}$ amplitudes as a function of pump fluence based on their fast Fourier-transform (FFT) peak heights obtained from the standard (isotropic) TR signals.  At high fluences the low-frequency tail of the asymmetrically broadened $A_{1g}$ peak overlapped and obscured the $E_g$ peak, however, which left considerable uncertainty in the quantitative evaluation of the latter mode.

In the present study we experimentally investigate coherent $E_g$ phonons in Bi under intense photoexcitation by employing an anisotropic detection scheme, in which the isotropic $A_{1g}$ contribution to the TR signals could in principle be cancelled.  To maximize the $E_g$ phonon signal we choose a bulk single crystal Bi whose crystalline axes are well specified and keep it at a low temperature of 11 K \cite{Ishioka2006, Misochko2006}.  For comparison we also examine the fluence-dependences of the $A_{1g}$ mode and photoexcited carriers in the standard (isotropic) detection scheme.   We find that the amplitudes of the two phonon modes exhibit strikingly contrasted fluence-dependences, and interpret the results in the context of dynamic coupling of the $E_g$ phonons with the $A_{1g}$ mode and with non-equilibrium photocarriers.
On the other hand, the initial phases of the two coherent phonons shift in parallel with increasing fluence, suggesting a fluence-dependent transition time from the ground state to the excited state.  

\section{Experimental}

The sample studied is a 1-mm thick bulk single crystal Bi with a (0001)-oriented polished surface in hexagonal notation (or (111) in cubic notation), which was purchased from MaTeck and is used without further treatment.  The crystal is mounted in a closed-cycle cryostat with its [11$\overline{2}$0] axis in vertical direction and is kept at 11 K.   

Single-color pump-probe reflectivity measurements are performed on the Bi crystal with an output of a regenerative amplifier with 120\,fs duration, 810 nm center wavelength (1.53 eV photon energy), and 100\,kHz repetition rate as the light source. A $f=100$-mm plano-convex lens focuses the linearly polarized pump and probe beams to the $\sim$80 and 40\,$\mu$m spots on the sample with incident angles of $<5^\circ$ and $15^\circ$ from the surface normal, respectively.  Incident pump fluence is adjusted between $F_\text{inc}=0.40$ and 28\,mJ/cm$^2$ by rotating a half-wave-plate before a fixed plate-type polarizer located in front of the focusing lens.  Pump beam is modulated with an optical chopper for lock-in detection.  
 
To examine the $E_g$ phonons we measure anisotropic transient reflectivity, $\Delta R_{eo}\equiv\Delta R_H-\Delta R_V$, by employing the incident probe polarized at $\sim45^\circ$ from horizontal ($H$) and by detecting the $H$ and vertical ($V$) polarization components of the reflected probe light with a pair of matched photodiode detectors.  
For comparison we also measure electronic and $A_{1g}$ responses in standard (isotropic) TR scheme, in which the pump-induced change in reflectivity $\Delta R$ is measured by detecting the probe light before and after reflection at the sample surface with a pair of matched photodetectors.  In both detections configurations, the signal from the detector pair is amplified with a current pre-amplifier and a lock-in amplifier. Time delay $t$ between the pump and probe pulses is scanned step by step with a translational stage (slow scan).  

\begin{figure*}
\includegraphics[width=0.85\textwidth]{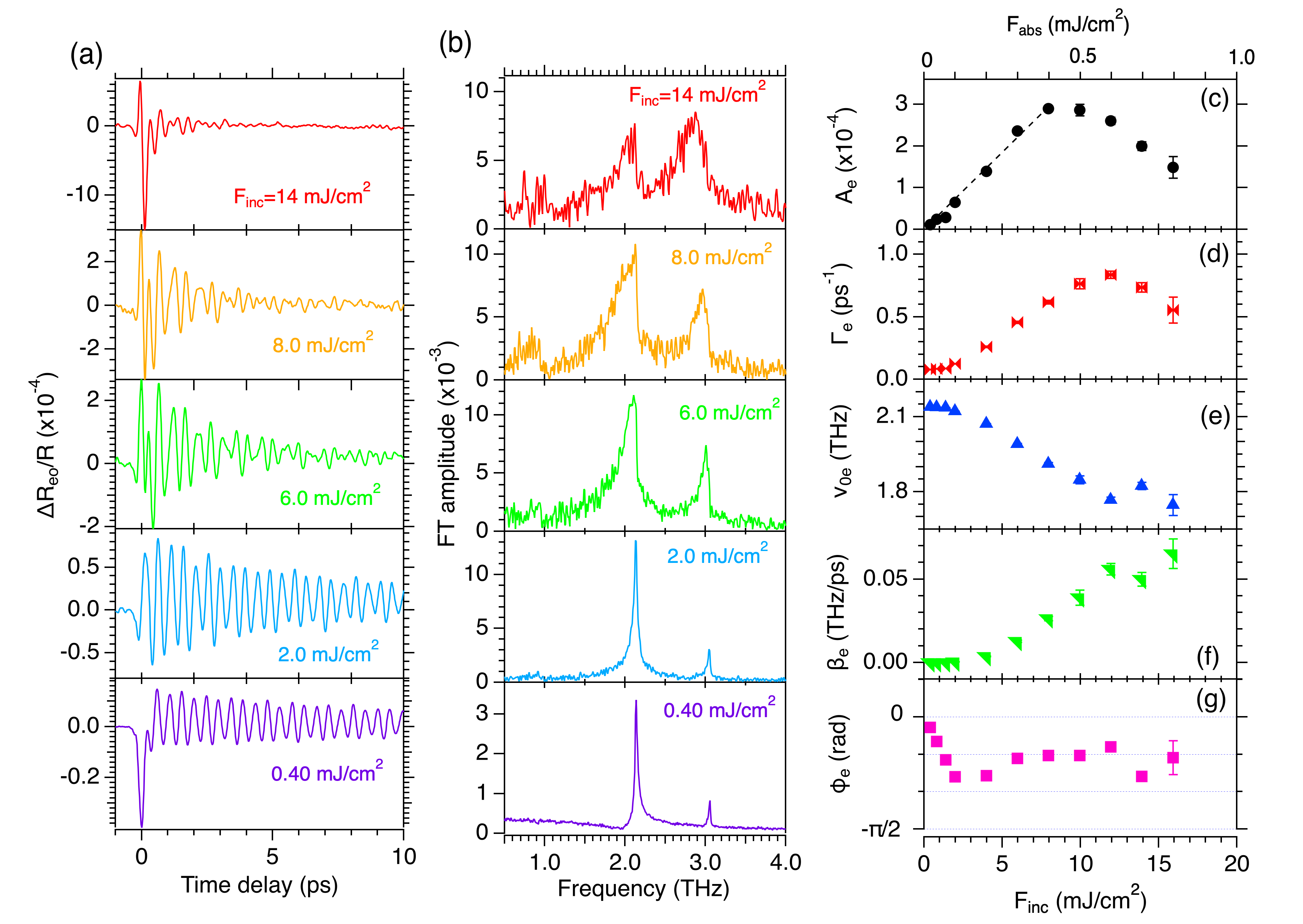}
\caption{\label{FEg} (a) Anisotropic TR signals of Bi(0001) surface obtained at 11 K with different pump fluences.  Pump light is polarized parallel to the [11$\overline{2}$0] axis.   
 (b) Fast Fourier transform (FFT) spectra of the oscillatory part of (a).  (c-g) Pump fluence depehdences of the amplitude (c), dephasing rate (d), initial frequency (e), frequency chirp (f), and initial phase (g) of the coherent $E_{g}$ phonon obtained by fitting the reflectivity signals in (a) to Eq.~(\ref{ddh}).  Dotted line in (c) indicates a linear fit in the low fluence regime.}
\end{figure*}

For a Bi crystal with \emph{moderate} coherent ionic displacements $Q_i$, photoexcited carrier density $N$, and lattice temperature $T_l$, the standard TR signal can be approximately expressed by the sum of their linear functions;
\begin{eqnarray}\label{Eq5}
\dfrac{\Delta R(t)}{R}=\dfrac{1}{R}\left[\sum_i\left(\dfrac{\partial R}{\partial Q_i}\right)Q_i(t)
+\left(\dfrac{\partial R}{\partial N}\right)N(t)\right.\nonumber\\
\left.+\left(\dfrac{\partial R}{\partial T_l}\right)\Delta T_l(t)\right],
\end{eqnarray}
with $i=y$ and $z$ denoting the displacements along the $E_g$ and $A_{1g}$ coordinates.  
In the anisotropic detection scheme the last two terms of Eq.~(\ref{Eq5}) as well as the $A_{1g}$ contribution to the first term are to be cancelled, since they are isotropic within the surface plane;  %
\begin{eqnarray}\label{Eq6}
\dfrac{\Delta R_{eo}(t)}{R}=\dfrac{1}{R}\left(\dfrac{\partial R}{\partial Q_y}\right)Q_y(t).
\end{eqnarray}
The anisotropic detection therefore enables us to monitor the $E_g$ phonon response exclusively.  
For a crystal far from the equilibrium under extremely intense photoexcitation, however, Eqs.~(\ref{Eq5},\ref{Eq6}) may no longer be adequate because of the contributions of the higher-order terms and/or because the crystal may be approaching photo-induced phase transition \cite{Cheng2022, Thiemann2024}.
 
\section{Results}

\subsection{Coherent $E_g$ Phonon}\label{SEg}

We first examine the $E_g$ phonon dynamics in the anisotropic detection scheme.  
Figure~\ref{FEg}(a) shows the anisotropic TR signals $\Delta R_{eo}$ at selected incident fluences $F_\text{inc}$.  Here the pump polarization is set to be parallel to the [11$\overline{2}$0] crystalline axis to maximize the $E_g$ phonon contribution, and the probe is polarized at $\sim45^\circ$ to it for the anisotropic detection \cite{Ishioka2006}.   At the minimum fluence of $F_\text{inc}=0.40$\,mJ/cm$^2$ the TR signal shows a negative spike at $t=0$, followed by a periodic modulation predominantly due to the $E_g$ phonon at $\sim$2\,THz.  Fast Fourier transform (FFT) spectrum in Fig.~\ref{FEg}(b) shows a small $A_{1g}$ peak at 3\,THz as well, however, due to the imperfect optical polarization in the experiments.  As the fluence is increased, the $E_g$ peak is broadened and redshifted, in agreement with the prior TR studies \cite{Misochko2006, Misochko2007}.  Meanwhile the $A_{1g}$ peak grows with fluence and eventually becomes higher than the $E_g$ peak. 
At the maximum fluence examined ($F_\text{inc}=16$\,mJ/cm$^2$) the baseline of the signal starts to exhibit large fluctuation in the middle of the slow scan of the time delay. 
We check if the exposure to the maximum fluence caused an irreversible damage on the sample surface by measuring the same spot measured at a low fluence. 
The obtained signal is as noiseless as that from a fresh spot,  the $E_g$ amplitude before and after the exposure is comparable, and the $A_{1g}$ mode becomes smaller than the $E_g$ again.  The results indicate that no significant irreversible damage such as melting was induced during the exposure, and that the enhancement of the $A_{1g}$ mode during the exposure is a transient phenomenon.
 
To quantitatively analyze the fluence-dependence of the coherent $E_g$ phonons we fit the oscillatory component of the time-domain TR signals to the sum of two damped harmonic functions:
\begin{eqnarray}\label{Eq9}
\label{ddh}
f(t)=A_{e} \exp(-\Gamma_{e} t)\sin[2\pi\nu_{e} t+\phi_{e}]\nonumber\\
+A_{a} \exp(-\Gamma_{a} t)\sin[2\pi\nu_{a} t+\phi_{a}],
\end{eqnarray}
where the subscripts $e$ and $a$ denote the $E_g$ and $A_{1g}$ modes.  For simplicity we assume linear chirps in the frequencies:
\begin{equation}\label{Eq8}
\nu_{i}=\nu_{0i}+\beta_{i} t,
\end{equation}
with $i$ denoting $e$ or $a$. This function can give excellent fits to the TR signals for $t>0.3$\,ps, 
whereas for $t<0.3$\,ps the fitting is somewhat poorer because of the large negative spike overlapping around $t=0$.  
Figure~\ref{FEg} (c-g) summarizes the $E_{g}$ phonon parameters obtained by the fitting as a function of incident pump fluence.
The initial frequency $\nu_{0e}$ redshifts from 2.1 to 1.8\,THz and the linear chirp $\beta_e$ increases from $<10^{-4}$ to 0.06\,THz/ps, whereas the dephasing rate $\Gamma_e$ increases from 0.08 to 0.8\,ps$^{-1}$,  with increasing fluence from 0.40 to 12\,mJ/cm$^2$.  

Surprisingly, the $E_g$ amplitude is found to increase almost linearly only up to $F_\text{inc}\simeq8$\,mJ/cm$^2$ and then to turn to an apparent decrease [Fig.~\ref{FEg}(c)].  We confirm that this behavior is independent of analysis method; the area under the $E_g$ peak in the FFT spectra similarly increases and then decreases with fluence. 
We note that the  fluence-dependence observed in the present study appears to be contradict the increase-and-saturation behavior reported in a previous TR study \cite{Misochko2006}.  This is because in the literature $E_g$ amplitude was estimated as the FFT peak height from the standard (isotropic) TR signals, in which much larger $A_{1g}$ signal obscured the $E_g$ contribution at high fluences. 

The initial phase of the $E_g$ phonon [Fig.~\ref{FEg}(g)] is found to be close to a sine function of time ($\phi_e=0$), as is expected for an excitation in the impulsive limit [Eq.~(\ref{Eq4})] with a $\delta$-function-like driving force.  With increasing fluence the phase first steeply decreases to $\phi_e\sim-\pi/4$ and then recovers partially to $\phi_e\sim-\pi/6$.  The deviation of the initial phase from zero indicates that the duration of the driving force can no longer be negligible compared to the phonon period.  We note that in the low fluence regime ($F_\text{inc}<6$\,mJ/cm$^2$), where we observe most prominent phase shift the fitting results are excellent [Fig.~S3 in SM], which ensures the reliable determination of the phase.  In the high fluence regime ($F_\text{inc}>6$\,mJ/cm$^2$) the fit leaves somewhat larger uncertainty, however, due to the faster dephasing of the $E_g$ oscillation as well as the larger contributions from the negative spike at $t=0$.  The poorer fitting may be the cause for the larger scattering in the initial phase in the high fluence regime.

To check the origin of the unconventional fluence-dependences of the phase shift we also perform similar anisotropic TR measurements using a longer pump pulse. 
We find $\phi_e$ to exhibit a qualitatively similar, if less pronounced, shift with fluence for a longer pulse duration, confirming that the phase shift is no artifact of the experiments.
 
\subsection{Electronic and Thermal Responses}\label{Snon}

\begin{figure}
\includegraphics[width=0.475\textwidth]{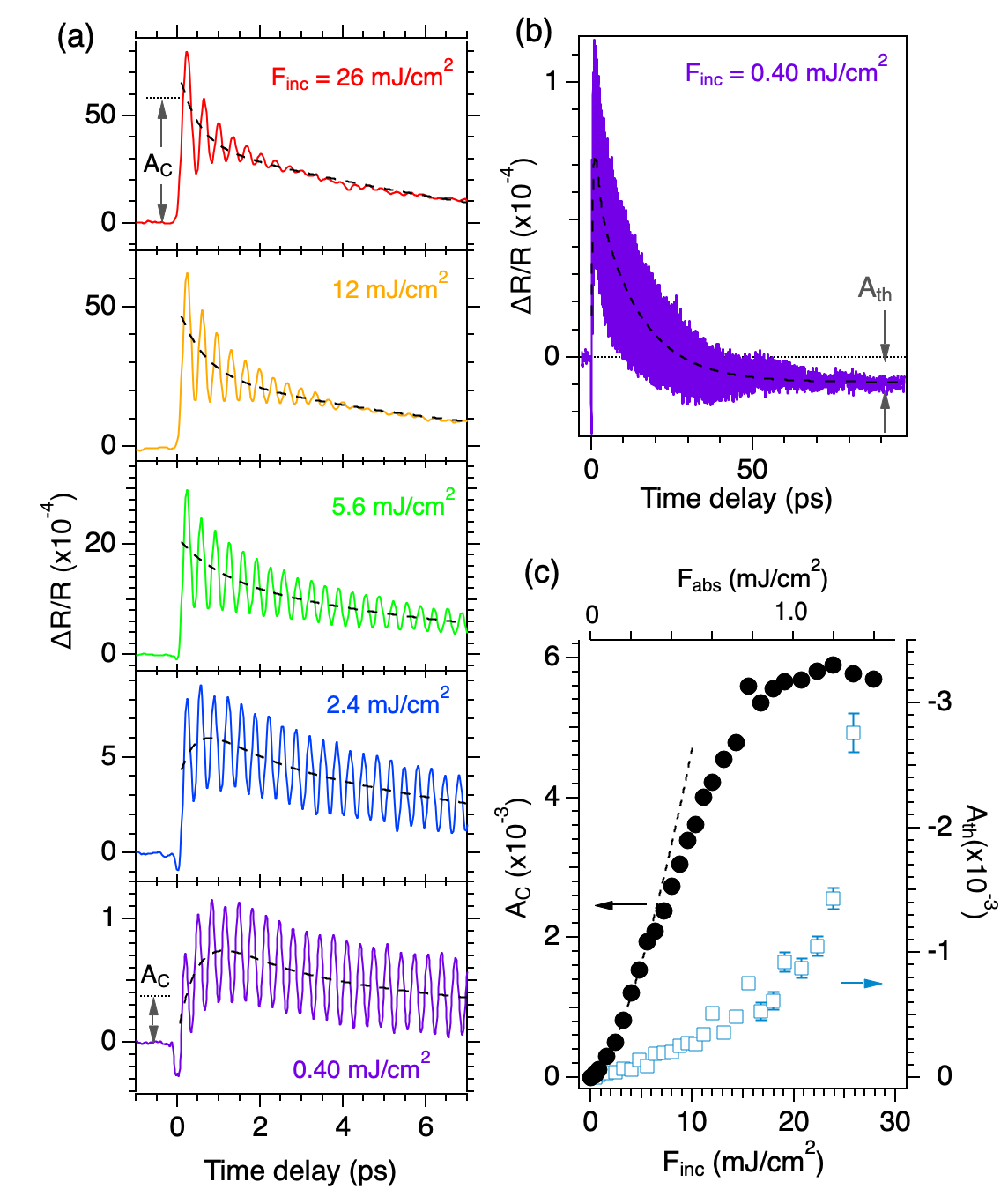}
\caption{\label{Fnon} (a,b) Standard TR signals of Bi (solid curves) obtained at 11 K at different pump fluences.  Broken curves represent the non-oscillatory response obtained by fitting to Eq.~(\ref{Eq7}). (c) Height $A_{C}$ of the non-oscillatory response at $t=0.2$\,ps (plotted to the left axis) and baseline $A_{th}$ at a long time delay (to the right axis) as a function of pump fluence.  Dotted curve represents the extrapolation of the fitting $A_{C}\propto F_\text{abs}^n$ in the low fluence regime.}
\end{figure}

For comparison we also examine the carrier and $A_{1g}$ phonon dynamics of the same Bi crystal sample excited under the same condition and detected in the standard TR scheme.  To minimize the $E_g$ contribution to the signals the pump and probe light polarizations are set at 45$^\circ$ and 0$^\circ$ to the [11$\overline{2}$0] axis, respectively \cite{Ishioka2006}.  The standard TR signals, shown in Fig.~\ref{Fnon}(a), feature a sizable non-oscillatory response of photoexcited carriers in addition to the oscillatory coherent phonon response.   The former can be fitted to a multi-exponential function on top of a baseline,  
\begin{equation}\label{Eq7}
g(t)=\Sigma_j A_j \exp (-t/\tau_j)+A_{th},
\end{equation}
whose results are indicated with broken curves in Fig.~\ref{Fnon}(a,b).  
At the lowest fluence examined ($F_\text{inc}=0.40$\,mJ/cm$^2$) the non-oscillatory component can be fitted to the sum of altogether three exponentials.  The time constant obtained for the rise, $\tau_\text{rise}=0.6$\,ps, is in agreement with that reported for the intervalley electron-phonon scatterings \cite{Timrov2012, Melnikov2013}.  The decay time constants are obtained to be $\tau_\text{fast}=0.85$\,ps and $\tau_\text{slow}=$13\,ps; the latter  is a few times slower than that of the electron-hole recombination reported in previous two-photon photoemission studies \cite{Timrov2012, Bronner2014}, possibly because of the thicker Bi crystal and the lower temperature employed in the present study.  At long time delays ($t>30$\,ps) the non-oscillatory component approaches a negative baseline $A_{th}$, as shown in Fig.~\ref{Fnon}(b).  The lattice temperature rise can be estimated from the value of $A_{th}$ to be $\Delta T_l<1$\,K by adopting the temperature-dependence of the reflectivity  \cite{Wu2007}:
\begin{equation}\label{temp}
\partial(\Delta R/R)/\partial T=-8\times10^{-5} \text{K}^{-1}.
\end{equation}

With increasing fluence the rise of the TR signal becomes faster in time, plausibly due to the larger contribution from the intraband scatterings among highly excited electrons and holes.  At $F_\text{inc}>4$\,mJ/cm$^2$ the rise is complete before the first maximum of the coherent oscillation ($\tau_\text{rise}\ll0.2$\,ps) and can no longer be fitted uniquely to an exponential function.  We therefore fit only the decay ($t>0.1$\,ps) to two exponentials.
In the following analyses we assume that the electrons and holes come to follow the Fermi-Dirac distribution before the first maximum of the oscillation at $t\simeq0.2$\,ps, and regard the non-oscillatory amplitude $A_C\equiv \Delta R_\text{non}(t=0.2\,\text{ps})/R$ as a semiquantitative measure for photoexcitated carrier density $N$, though this assumption may no longer hold at high fluences.
We find that $A_{C}$ grows first superlinearly ($A_{C}\propto F_\text{abs}^{n}$ with $n=1.5$) up to $F_\text{inc}\simeq8$\,mJ/cm$^2$ (dotted curve in the figure), then turns to a linear increase until $A_{C}$ reaches a saturation at $\gtrsim20$\,mJ/cm$^2$, as shown with filled symbols in Fig.~\ref{Fnon}(d).  
Further increase in the fluence eventually leads to an emergence of large noise in the TR signal at $F_\text{inc}=28$\,mJ/cm$^2$. 
When the fluence is lowered after the exposure to the maximum fluence, we obtain a noiseless TR signal with distinct coherent phonon oscillation again.

\begin{figure}
\includegraphics[width=0.475\textwidth]{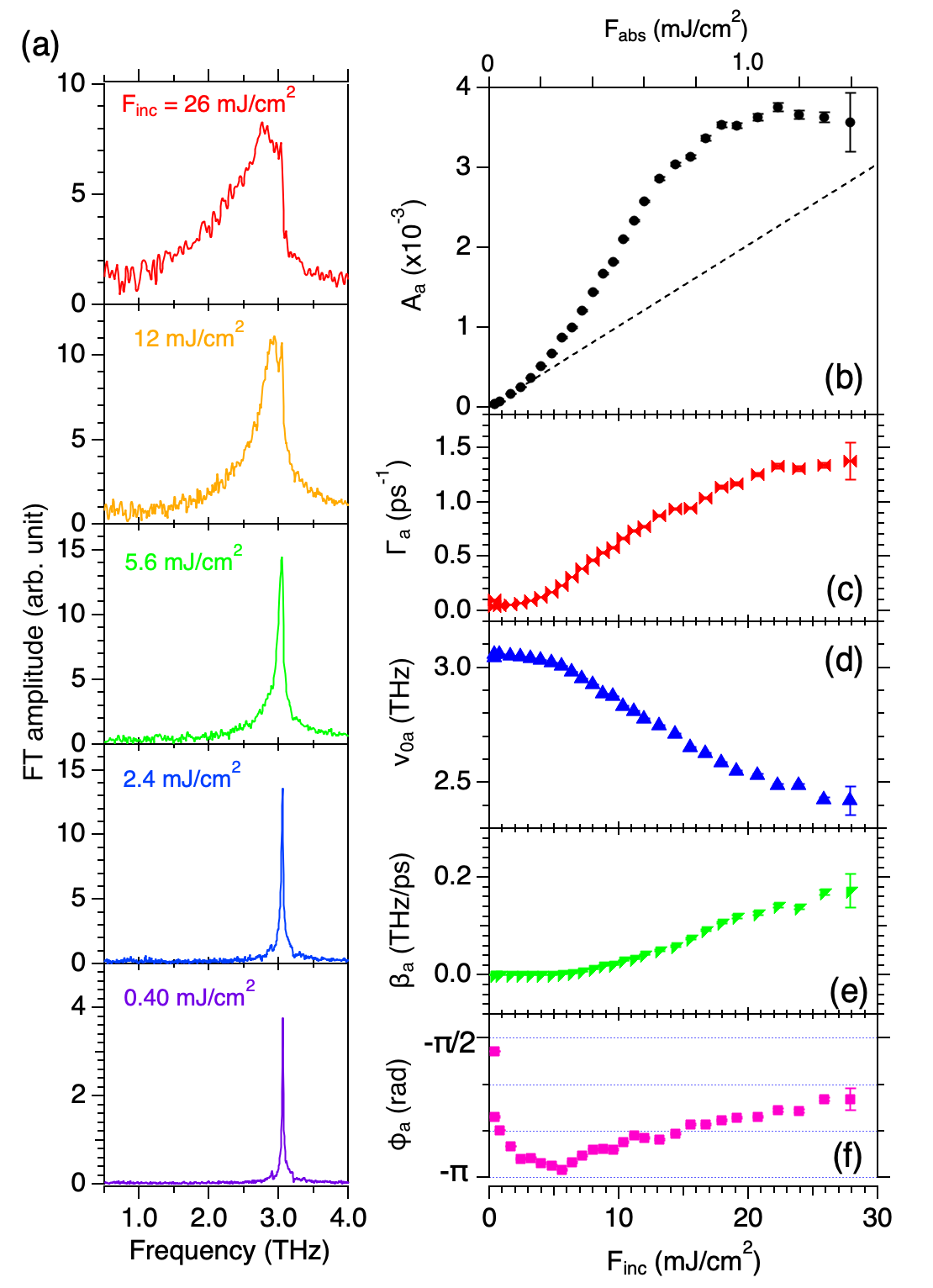}
\caption{\label{FA1g} (a) FFT spectra of the oscillatory part of TR signals shown in Fig.~\ref{Fnon}(a).  (b-f) Fluence-dependences of the amplitude (b), dephasing rate (c), initial frequency (d), frequency chirp (e), and Initial phase (f) of the coherent $A_{1g}$ phonon, obtained by fitting the oscillatory reflectivity signals to the second term of Eq.~(\ref{Eq6}).  Dotted line in (b) represents the extrapolation of a linear fitting in the low fluence regime ($F_\text{inc}<3$\,mJ/cm$^2$). }
\end{figure}

\begin{figure*}
\includegraphics[width=0.975\textwidth]{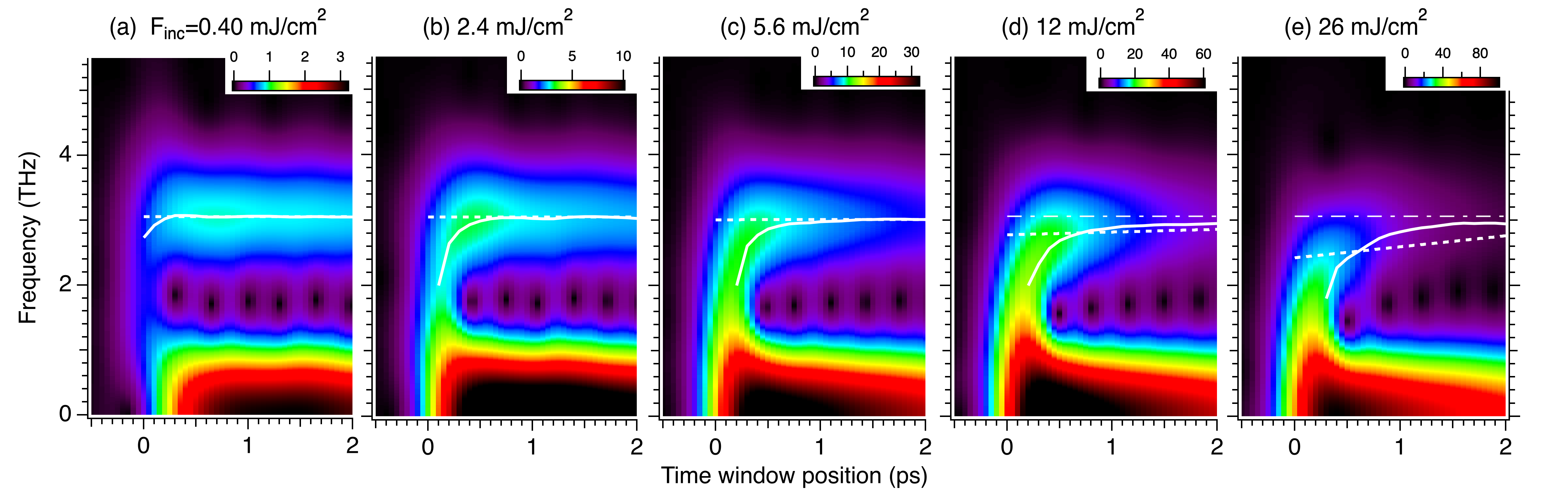}
\caption{\label{TWFT} False-color plots of time-windowed FFT amplitude as a function of time window position and frequency obtained at different fluences.  A Gaussian time window of 330-fs half width at half maximum is used.  Solid curves represent the peak positions within the frequency range of 2 $-$ 3.5 THz.  Broken lines reproduce the linearly chirped $A_{1g}$ frequency $\nu_a=\nu_{a0}+\beta_a t$ obtained from time-domain fitting.  Chained lines in (d,e) represent the intrinsic $A_{1g}$ frequency.}
\end{figure*}

The saturation of $A_{C}$ at $F_\text{inc}>20$\,mJ/cm$^2$ may be interpreted as the saturation of linear optical absorption, which is mainly responsible to the early time response of the single-color TR signals.  To assess the possibility we estimate the carrier density in the following.  At $t=0$ the photoexcited carrier density $N$ should have a depth distribution described by an exponential function of the distance from the surface $z$: $N(t=0, z)=N_0\exp(-\alpha z)$. A recent TR study on thin ($\ll1\,\mu$m) Bi films reported that the depth distribution becomes homogenized within 150 fs and its density at the surface reduces to 1/3 - 1/5 depending on the film thickness \cite{Thiemann2022}.   Such ultrafast homogenization of photocarriers would be less likely in the bulk crystal, and for simplicity we assume the incident pump light with photon energy $h\omega$=1.53 eV to be absorbed uniformly and completely within the optical penetration depth $1/\alpha=14.7$\,nm:
\begin{eqnarray}
N(t\simeq0)=\left\{
\begin{array}{ll}
(1-R)F_\text{inc}\alpha/(h\omega) &\text{for}\quad 0<z<1/\alpha,\\
0 &\text{for}\quad z>1/\alpha,
\end{array}
\right.
\end{eqnarray}
which would give an upper limit to the carrier density at the surface.  As for the reflectivity at 11\,K we tentatively estimate $R=0.95$ by extrapolating the reflectivity reported for higher temperatures \cite{Cardona1964, Kubota2023}. 
The absorbed fluence $F_\text{abs}\equiv(1-R)F_\text{inc}$ obtained by using this reflectivity value is plotted on the top axes of Figs.~\ref{FEg}(c-g), \ref{Fnon}(c) and \ref{FA1g}(b-f).  This choice is justified by the quantitative agreement of the fluence-dependence of the $A_{1g}$ frequency 
with that reported for a 197-nm thick Bi film at room temperature \cite{Thiemann2022}.
At $F_\text{inc}=20$\,mJ/cm$^2$, at which $A_{C}$ reaches a saturation, we would obtain $N=2.8\times10^{21}$\,cm$^{-3}$.  This density would correspond to 2\% of the valence electrons of Bi:
\begin{equation}
N_v=5 N_A w /M =1.4\times10^{23} \text{cm}^{-3}, 
\end{equation}
with 5 being the number of valence electrons per Bi atom, $N_A$, the Avogadro constant, $w=9.747$ g/cm$^3$, the density of Bi, and $M=209$, the atomic number.  At such a high carrier density the linear absorption would possibly be saturated and lead to a saturation of $A_C$. 

At long time delays ($t\gg10$\,ps) we can obtain the baseline $A_{th}$ by fitting the TR signals to Eq.~(\ref{Eq7})  and estimate the equilibrium lattice temperature using Eq.~(\ref{temp}).  $A_{th}$ increases almost linearly up to $F_\text{inc}=20$\,mJ/cm$^2$ and then turns to a steeper increase with fluence, as plotted with open symbols in  Fig.~\ref{Fnon}(c).  At the maximum fluence examined, the lattice temperature rise is estimated to be $\Delta T_l\simeq35$\,K.

\subsection{Coherent $A_{1g}$ Phonon}\label{SA1g}

The $A_{1g}$ phonon response can be extracted  from the TR signals in Fig.~\ref{Fnon}(a) by subtracting the non-oscillatory component (broken curves).   The FFT spectra shown in Fig.~\ref{FA1g}(a) features only the $A_{1g}$ peak at 3\,THz, confirming that the $E_g$ contribution is minimized with the selected pump and probe light polarizations.  
We accordingly fit the oscillatory oscillatory signals to only the second term of Eq.~(\ref{Eq9}), either over the entire time range or over the first three cycles of the oscillations as was done in some of the previous studies.  
Although neither reproduces the experimental oscillations perfectly at all the pump fluences, the $A_{1g}$ phonon parameters obtained from the two fittings are in reasonable agreement.  We therefore discuss on only the phonon parameters obtained from the fitting over the entire time window, which are summarized in Fig.~\ref{FA1g}(b-f).   

The $A_{1g}$ amplitude increases linearly in the low fluence regime ($F_\text{inc}<3$\,mJ/cm$^2$), as indicated with a dotted line in Fig.~\ref{FA1g}(a), but grows superlinearly ($A_a\propto F_\text{inc}^{1.5}$) in the intermediate fluence regime ($F_\text{inc}=3-15$\,mJ/cm$^2$). 
Further increase in the fluence leads to a saturation of the amplitude in the high fluence regime ($F_\text{inc}>20$\,mJ/cm$^2$) and eventually to an emergence of large noise in the TR signal during the time delay scan. 
When the fluence is lowered after the exposure to the maximum fluence, we obtain coherent $A_{1g}$ oscillation whose initial amplitude is almost as large as that before the exposure, but the dephasing is much faster, as shown in Fig.~S2(b,c).  The comparison indicates a small but irreversible change in the laser-irradiated crystal, such as slight damage at the surface, but no sign of complete melting or phase transition.

The observed fluence-dependence of the $A_{1g}$ amplitude is in quantitative agreement with a prior TR study performed in a similar condition (on a 1-mm thick Bi single crystal at 5\,K) \cite{Misochko2009}. 
The behavior is also in rough, though not perfect, agreement with that of the electronic response $A_{C}$ [filled symbols in Fig.~\ref{Fnon}(c)], suggesting that the saturation of $A_{1g}$ amplitude has the same origin as that of $A_C$.  On the other hand, it is in striking contrast to the fluence-dependences of the $E_g$ amplitude [Fig.~\ref{FEg}(c)], which turns to decrease already at $F_\text{inc}\simeq10$\,mJ/cm$^2$ while the $A_{1g}$ amplitude is still growing superlinearly.  The contrast \emph{excludes} the possibility that the $E_g$ amplitude decreases because the entire crystalline lattice is becoming unstable, by approaching the high-symmetry phase for example.

The initial phase $\phi_a$ is close to $-\pi/2$ at the minimum fluence of $F_\text{inc}=0.4$\,mJ/cm$^2$, as expected for coherent phonons excited in the displacive limit (Eq.~(\ref{Eq2})) with a Heaviside-step-function-like driving force.  With increasing fluence the initial phase first shifts steeply down to $\phi_a\simeq-\pi$ and then gradually recovers toward $\phi_a=-2\pi/3$.  We note that fitting results are excellent throughout the entire fluence range examined, 
ensuring the reliable determination of the initial phase.   We also observe a consistent phase shift with fluence in the standard TR measurements using a longer pump pulse. 
In the DECP model the initial phase can deviate from $\pm\pi$ by introducing a finite decay time for the driving force that is assumed to be proportional to the carrier density $N$ or electronic temperature $T_e$ \cite{Zeiger1992}.  
We notice, however, that the initial phase of the $E_g$ phonon [Fig.~\ref{FEg}(g)] also shifts with fluence, almost in parallel to that of the $A_{1g}$. 
The parallel trends of the two modes could be better explained by a fluence-dependent \emph{rise} time in the driving forces, rather than a decay time, since the impulsive ($\delta$-function-like) driving force for the $E_g$ phonon is less likely to be affected by the decay in $N$ or $T_e$.

The rising of carrier-phonon coupling that is responsible to the coherent phonon generation can be visualized by performing time-windowed Fourier transform on the whole TR signals containing both the electronic and phononic responses [colored curves in Fig.~\ref{Fnon}(a)].  The results obtained with a Gaussian time window of half width $\Delta t=0.33$\,ps, which allows us the best balance between the temporal and frequency resolutions ($\Delta \nu\simeq0.5$\,THz), are presented as false-color plots in Fig.~\ref{TWFT}.  At the lowest fluence [Fig.~\ref{TWFT}(a)] the phononic response at $\sim$3\,THz and the electronic response at $\lesssim1$\,THz are well separated from each other; they both emerge within the time window width used in the analysis.  With increasing fluence the electronic response acquires a higher-frequency component at early time delays, which interconnects it with the phononic response in the first fraction of ps.   At later time delays the coupled electron-phonon response [solid curves in Fig.~\ref{TWFT}] approaches the intrinsic $A_{1g}$ phonon frequency [chained curves in the same figure].  The time scale for this transient blueshift is comparable to the time window width at the minimum fluence ($F_\text{inc}=0.4$\,mJ/cm$^2$) but becomes slower with increasing fluence up to $\gtrsim$1\,ps at the maximum fluence.  We interpret the time scale in terms of the fluence-dependent transition time from the ground-state PES to the excited-state PES, as was predicted for the $A_1$ phonon of tellurium, which is also associate with the Peierls stability, by recent time-dependent DFT simulations \cite{Ning2022}.
Indeed we find a notable discrepancy between the transient shift of the coupled electron-phonon response and that of the $A_{1g}$ phonons driven by the coupling [dotted lines in Fig.~\ref{TWFT}] at the early times, the latter of which is estimated from the time-domain fitting to Eq.~(\ref{Eq9}) at respective excitation density.  The discrepancy can be regarded as the manifest of the excited-state PES still shifting and deforming towards a new equilibrium.  The two frequencies eventually coincide after the system reaches the equilibrium on sub-picosecond time scale in the low to medium fluence regime.  In high fluence regime ($F_\text{inc}>15$\,mJ/cm$^2$), however, the two frequencies no longer coincide on reasonable time scale, implying the failure of assuming a linear frequency chirp in this regime.

\section{discussion}

\begin{figure}
\includegraphics[width=0.475\textwidth]{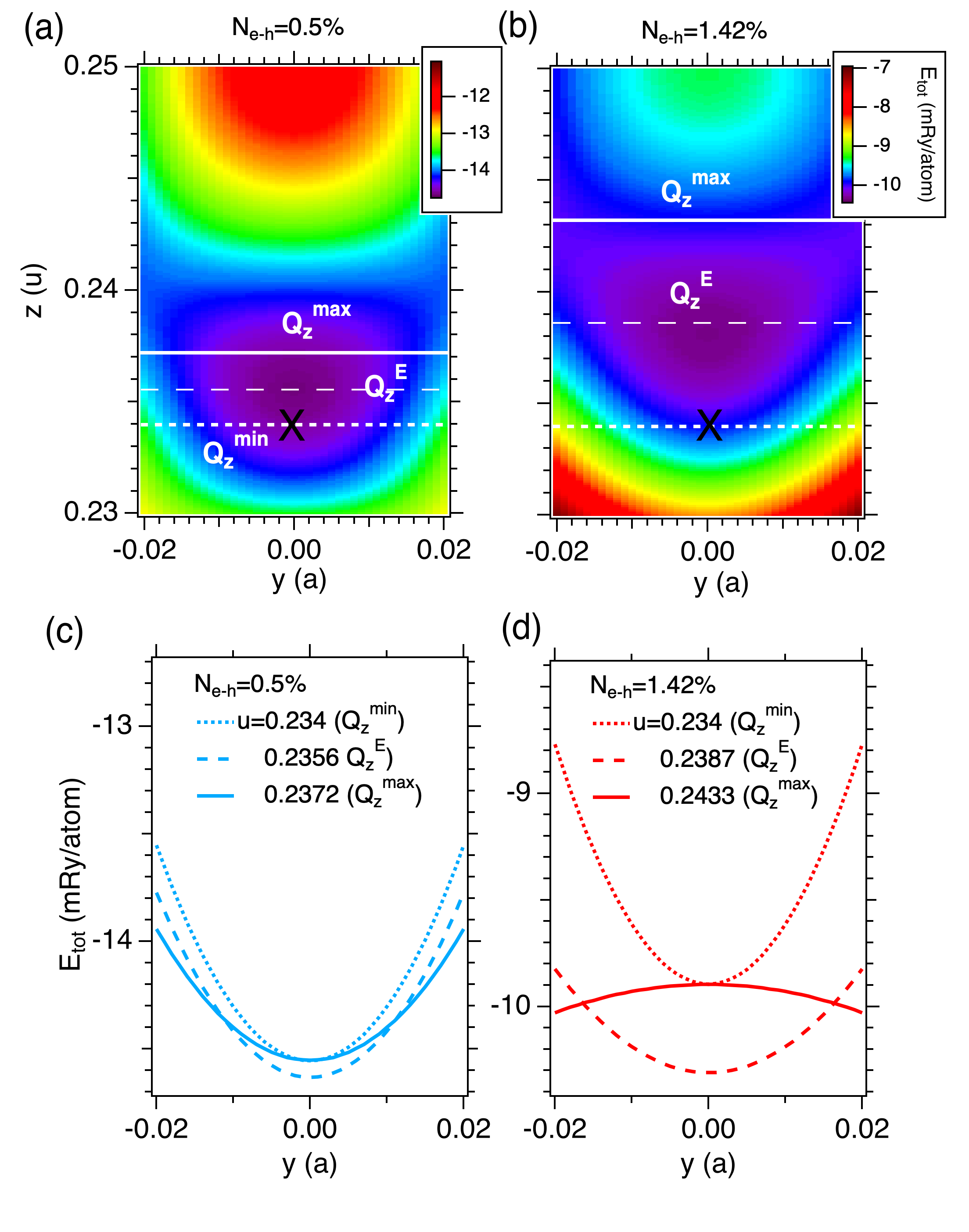}
\caption{\label{FPES} (a,b) PESs as functions of the $A_{1g}$ and $E_g$ coordinates for excited electronic states at photoexcitation of 0.5\% (a) and 1.42\% (b) of valence electrons, as reproduced by using the parameters obtained in the DFT simulations \cite{Zijlstra2006}.  The $E_g$ ($y$) and $A_{1g}$ ($z$) coordinates are given in units of the hexagonal lattice parameter $a$ and the internal displacement $u$.  Cross represent the ground-state minimum at $z=Q_z^\text{min}=0.234$ and $y=0$.
Broken and dotted lines the equilibrium on the excited state $Q_z^E$  and the maximum displacement $Q_z^\text{max}$ in the $A_{1g}$ coordinate.  (c,d) Slices of the two-dimensional PES along the $E_g$ coordinate at different values of $u$ at photoexcitation of 0.5\% (c) and 1.42\% (d) of valence electrons.
}
\end{figure}

We now discuss the origin of the unconventional fluence-dependence of the $E_g$ phonon amplitude [Fig.~\ref{FEg}c], which is in stark contrast to the behavior of the $A_{1g}$ amplitude [Fig.~\ref{FA1g}b].

A prior DFT study \cite{Zijlstra2006} calculated excited-state PESs as functions of $E_g$ and $A_{1g}$ coordinates. 
Close-ups of the reported two-dimensional (2D) PESs around the ground-state equilibrium ($z=Q_z^\text{min}=0.234$ and $y=0$, where $z$ and $y$ are represented in unit of $u$ and $a$) are reconstructed in Fig.~\ref{FPES}(a,b) for two selected densities of photoexcited electron-hole pairs $N_{e-h}$.  
In the $z$ coordinate the equilibrium $Q_z^E$ shifts with photoexcitation, and Bi ions oscillate between $Q_z^\text{min}$ and $Q_z^\text{max}=2Q_z^E-Q_z^\text{min}$ via the DECP mechanism, as also shown schematically in Fig.~\ref{rhombo}(b).
In the $y$ coordinate the ions oscillate around $y=0$ via ISRS mechanism, whose amplitude is  supposed to be an-order-of-magnitude smaller than the $A_{1g}$ \cite{Johnson2013}.

At a low excitation density, where the PES can be approximated to be harmonic along both coordinates, one could assume that the $E_g$ and $A_{1g}$ oscillations are independent of each other.  Even in the case of $N_{e-h}=0.5\%$ of valence electrons, however, the PES slice along the $y$ coordinate, shown in Fig.~\ref{FPES}(c), is not perfectly independent of the $A_{1g}$ displacement but receives a small disturbance as a function of $z$.
With increasing $N_{e-h}$ the equilibrium $Q_z^E$ shifts toward the central barrier in the $z$ coordinate, while the barrier between the double well becomes lower, as shown 
in Fig.~\ref{rhombo}. This, together with the increasing $A_{1g}$ displacement, introduces a significant deformation in the PES slice along the $y$ coordinate while $z$ varies from $Q_z^\text{min}$ to $Q_z^\text{max}$, i.e., within a half cycle of the $A_{1g}$ oscillation.
At $N_{e-h}=1.42\%$, the PES along the $y$ coordinate suffers so considerable deformation that 
the curvature of the PES slice becomes negative when the ion reaches $Q_z^\text{max}$, as shown in Fig.~\ref{FPES}(d).  We infer that this deformation of the PES would lead to a quick loss of the vibrational coherence of the $E_g$ mode within a single cycle of the $A_{1g}$ oscillation, and thereby to an effective suppression of coherent $E_g$ phonons at high fluences as observed in Fig.~\ref{FEg}.  

Further increase in the excitation density above 2\% would transform the double-well PES in the $z$ coordinate to a single well, as illustrated in Fig.~\ref{rhombo}(c), according to another DFT simulation \cite{Murray2005}.  A single-shot TR study \cite{Teitelbaum2018} reported a gradual decrease in the $A_{1g}$ oscillation amplitude under intense photoexcitation and complete disappearance at  $F_\text{inc}>10$\,mJ/cm$^2$ ($F_\text{abs}>3$\,mJ/cm$^2$) for a 275-nm thick Bi film at room temperature.  The disappearance was interpreted as a result of photoinduced transition to the theoretically predicted high-symmetry phase.   
 A trXRD study on a 50-nm thick Bi film at room temperature reported a similar disappearance of the oscillation at $F_\text{abs}>3$\,mJ/cm$^2$ \cite{Lu2009}.  In the present study we did not observe such complete disappearance of the oscillation, however, because we basically stayed below the threshold fluence.  Since the $A_{1g}$ mode exhibited larger frequency redshift and faster dephasing  for thinner Bi films due to higher carrier concentration confined  \cite{Shin2015,  Teitelbaum2018, Jnawali2021, Thiemann2022, Thiemann2024} it would be reasonable to assume that even higher absorbed fluence would be required to induce a phase transition in a thick bulk crystal than in a thin film.    Instead we observed large noise in the TR signal starting to appear during the scan of time delay
, which is indicative of partial damage of the crystal surface as a result of continuous heating during the repetitive excitation at 100~kHz at a high fluence.  We speculate that the lattice cooling at the surface of a 1-mm thick Bi crystal could be less efficient than in a sub-$\mu$m-thick film on a substrate due to the low thermal conductivity of Bi \cite{Gallo1963} than that of the substrate.

\section{Conclusion}

Ultrafast dynamics of coherent $E_g$ phonons of bulk single crystal Bi was investigated under intense photoexcitation  at low temperature.  With increasing pump fluence the $E_{g}$ amplitude reached its maximum and turned to a decrease at a significantly lower fluence than the $A_{1g}$ amplitude became saturated.  The contrasted behaviors were explained in terms of the strong coupling of the $E_g$ oscillation with the large-amplitude $A_{1g}$ oscillation on the highly excited electronic state, which could lead to suppression of the former oscillation via dynamic deformation of the PES.  By contrast, the fluence-dependence of the $A_{1g}$ phonons are dominated by the coupling with photoexcited carriers rather than that with the $E_g$ phonon.  The phase shifts of the two phonon modes with fluence could be an indication of carrier density-dependent transition time from the ground-state PES to the excited-state one.  

\begin{acknowledgments}
KI thanks Dr. Yuya Kubota and Dr. Yasushi Shinohara for their valuable discussion.
\end{acknowledgments}

\bibliography{Bi_ref}

\end{document}